\documentclass[review,12pt]{elsarticle}

\usepackage[utf8]{inputenc}
\usepackage{graphicx}
\usepackage{amsmath}
\usepackage{amssymb}
\usepackage{booktabs}
\usepackage[margin = 0.8in]{geometry}
\usepackage{float}

\journal{a journal}

\begin{document}

\begin{frontmatter}



\title{Deep neural networks based predictive-generative framework for designing composite materials}

\author[add1]{Ashank}
\address[add1]{Department of Mechanical Engineering, Indian Institute of Technology Ropar, Rupnagar-140001, India}
\author[add1]{Soumen Chakravarty}
\author[add1]{Pranshu Garg}
\author[add1]{Ankit Kumar}
\author[add1]{Manish Agrawal}
\author[add1]{Prabhat K. Agnihotri}
% \affiliation{organization={},%Department and Organization

\begin{abstract}
Designing composite materials as per the application requirements is fundamentally a challenging and time consuming task. Here we report the development of a deep neural network based computational framework capable of solving the forward (predictive) as well as inverse (generative) design problem. The predictor model is based on the popular convolution neural network architecture and trained with the help of finite element simulations. Further, the developed property predictor model is used as a feedback mechanism in the neural network based generator model.  The proposed predictive-generative model can be used to obtain the micro-structure for maximization of particular elastic properties as well as for specified elastic constants. One of the major hurdle for deployment of the deep learning techniques in composite material design is the intensive computational resources required to generate the training data sets. To this end, a novel data augmentation scheme is presented. The application of data augmentation scheme results in significant saving of computational resources in the training phase. The proposed data augmentation approach is general and can be used in any setting involving the periodic micro-structures. The efficacy of the predictive-generative model is demonstrated through various examples. It is envisaged that the developed model will significantly reduce the cost and time associated with the composite material designing process for advanced applications. 
\end{abstract}


% \begin{highlights}
% \end{highlights}

\begin{keyword}
Composite Material, Microstructure, Machine Learning, Neural Network, Data Augmentation.


\end{keyword}

\end{frontmatter}

%\linenumbers

\section{Introduction}
\label{Intro}
The need for lighter yet stronger and multifunctional materials has put emphasis on the development of tailored composites for aerospace, biomedical, and electronics industries. The conventional approaches of designing the composite materials is expensive, tedious, and trial-and-error based  that requires skilled domain expertise, intuition and a bit of luck. Commonly, the experimental techniques and/or finite element method (FEM) based routes are employed to optimize the composition and configuration of composite materials. While experiment based make and break route is quite costly, computation techniques like FEM  are  computationally expensive. A large number of experiments and/or numerical iterations might be required to obtain the optimum deign among the possible permutations/combinations. Thus, one has to put a lot of time and effort into designing the most optimal composite material through these routes. The machine learning (ML) based approach has a huge potential in accelerating the process of designing optimal composite materials and saving a lot of time and resources \cite{p8}.  With the rapid development of machine learning techniques and artificial intelligence (AI) approaches, many industries have seen improvement in their existing processes \cite{p1}. ML has been useful in building solutions for speech detection, spam filtering,  search engines,  disease,  and drug discovery \cite{p2,p3}.  Machine  learning  has  been  employed  to  discover  new  materials  with  optimal  designs  and  to predict the properties of different mechanical and structural systems \cite{p4,p5,p6,p7}. In 2020, Kollman \textit{et al}., developed a deep learning (DL) model based on convolutional neural networks (CNN) to predict the optimal metamaterial designs \cite{x1}. In another work, Baekjun \textit{et al}., built an artificial neural network to generate the optimal design of the porous materials given the required physical properties \cite{x3}. Thus, the ML techniques have provided an efficient mechanism to learn the material structure and property relationships based on the simulation and the experimental data.
       %and with enough training dataset and relevant machine learning techniques,  an  ML  model  can  be  trained  to  predict  the  properties  of  the  given  composite  material design as well and in this manuscript we have made an attempt to design the composite materials in a systematic and computationally efficient manner using the deep learning techniques

Fundamentally there are two scenarios which are required to be addressed while designing a composite material. Firstly, given the microstructure of the composite material  constituent properties, how to predict the composite material’s macroscopic properties (forward or predictive model) and secondly, given the final macroscopic properties of the composite, what will be the optimal microstructure of the composite material (inverse or generative model). While the ML-based approach to predict the composite material properties is in a nascent stage, attempts have been made by  multiple  research  groups in the  last few  years \cite{p9}. Most  of the existing  work  is  on  training  ML  models  to  predict  homogenised  properties \cite{n1,p10,p11,p12,p13} for the given microstructure of the composite material. In these efforts, the microstructure is represented as an image and the data obtained from the FEM simulations is used to train the neural network (NN). The FEM analysis is performed on the representative volume element (RVE) with periodic boundary condition to obtain the homogenized properties. Gu et al., in 2017 \cite{n1}, worked on designing composites using deep learning algorithms and demonstrated the ability of these algorithms to predict the mechanical properties of a composite such as toughness and strength. Yang \textit{et al}., \cite{p13} built a convolutional neural network to predict the properties of the composite materials beyond the elastic limits and demonstrated that ML has the potential to accelerate the process of composite design optimization. In another work, Pathan \textit{et al}.,\cite{p12} worked on predicting the homogenized mechanical properties of unidirectional fiber composites using supervised machine learning by employing a gradient-boosted tree regression model with 10-fold cross-validation strategy. In another attempt to predict the composite material properties using a convolutional neural network(CNN), \cite{p11} developed a neural network model that captures the elastic modulus, strength, and toughness of composites with high accuracy. They also integrated the CNN model with a genetic algorithm optimizer to search for the optimal microstructural design.
    
As shown by various researchers cited above, the property predictions through ML models are computationally efficient than using FEM analysis. Still the usefulness of these property prediction models are limited. The problem of obtaining the microstructure for a given set of macroscopic properties is far more crucial and complex to solve and its impact in accelerating the process of finding the optimal design will be significant. The method to obtain the micro-structure design for a specified properties is known as inverse homogenisation technique.  Traditionally, gradient based schemes like SIMP~\cite{x2} have been also applied towards solving the inverse homogenization problem~\cite{sigmund1994materials,huang2011topological}. In the context of ML based model, the literature for obtaining the composite microstructural designs is very limited. In the  literature, researchers have worked with two distinctly different approaches (i) The deep learning model has been trained for the optimized design obtained from the methods like SIMP for various loading and other design parameters. Once trained, the network can then be applied to obtain the optimum design for a new given condition~\cite{x3}. (ii) In another approach, a predictive-generative deep neural network to obtain the microstructure with high toughness is proposed~\cite{chen2020generative}.  The generator is based on gradient based scheme and it is connected to a  surrogate predictor model based on the deep neural network. Since, the generator is based on gradient scheme the intermediary densities are also allowed in the optimization framework. The major limitation of the first approach is that it relies on the existing optimum designs from other methods. The second approach is much more general and holds greater potential. But, the challenge in the second approach is that the training of the surrogate predictor model requires a large amount of training dataset. The training dataset is generated using FEM analysis, which is a computationally expensive task, and evidently, the process of data generation is slow and tedious. The other limitation of the current methodology is that requirement of the binary design is not imposed, and thus  post-processing is required to obtain the binary designs. 
    
In this work, we adopt the predictive-generative model approach to find the composite material micro-structure for achieving desirable homogenized elastic material properties. The proposed predictive-generative model can be used to obtain the micro-structure for maximization of particular elastic properties or to find tailored micro-structure having specified elastic constants. In both of these settings, additional requirements like specified fiber volume or the minimum fiber fraction can be imposed. To the best of authors knowledge, this is the first time that the optimum micro-structure is predicted for the specified elastic properties of composite material using ML based computational model. As described above, the major limitation of the existing approach is the high computational resources requirement during the training phase of the predictor model. To overcome this, we propose a novel two-step data augmentation scheme which manifolds our training data significantly without any additional computational efforts. The data augmentation approach increases the data set size by a factor of $8\times N\times N$ [where $N$ is the number of a grids on the edge of a representative volume element (RVE)] and thus leads to significant reduction in the computational resources requirements. The data augmentation technique exploits the periodic nature of the  RVE, and thus can be readily deployed in any applications involving the periodic microstructure.  Lastly, to ensure that obtained designs have very minimal presence of gray areas, the additional constraint is also imposed in the optimization framework. Thus directly binary designs can be achieved without any need for further post processing. The efficacy and the robustness of the proposed framework has been shown through various numerical simulations. To demonstrate the generality of the algorithm, simulation with various settings like given volume fraction constraints, minimum fibers are also presented.
    
The  remainder  of  the  paper  is  organized  as  follows:   Section  \ref{FEM}  presents the  of finite element analysis  on the RVE to calculate the effective macroscopic properties of the composite materials.  Section~\ref{CNN} discusses the convolution neural network (CNN) and how it can be used as a surrogate property prediction model.  Section \ref{DATA} explains the novel approach of data augmentation scheme developed for generating the training data of the property prediction model and discusses how these techniques result in significant saving of computational resources for training data generation.  Section \ref{GEN}  explains  the  architecture  and  working  of  the  inverse  generator  model  developed  to  design the composite microstructures.  In Section~\ref{RESULT}, the results obtained for the predictor model  and  the  inverse  generator  model are discussed and  finally the key findings of the work are summarized in Section \ref{CONCLUSION} .
\section{ Methodology}
    \subsection{ Theory and Finite Element Modelling} \label{FEM}
\begin{figure}[h]
\centering
\includegraphics[scale =1 ]{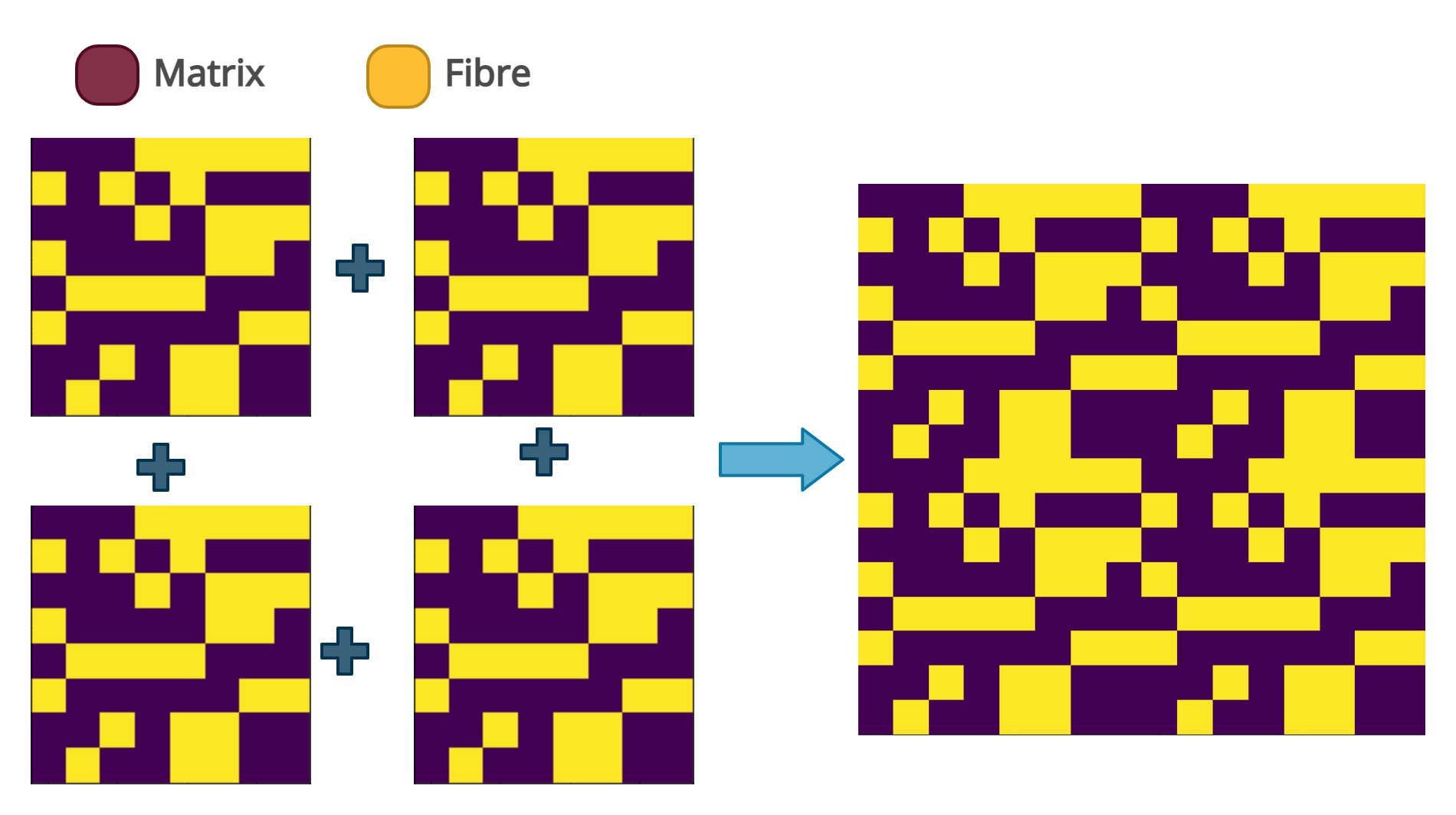}
\caption{Periodic stacking of RVE.}
\label{Fig.1}
\end{figure}

\begin{figure}[H]
\includegraphics[scale = 2]{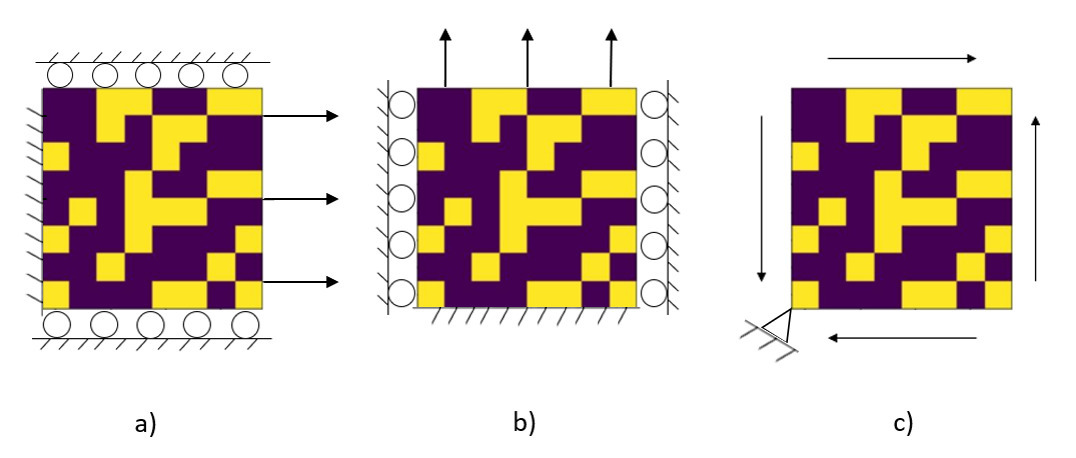}
\caption{Applied load schematic a) Longitudinal tension b) Transverse tension c) Pure shear}
\label{Fig.2}
\end{figure}

\begin{figure}[H]
\centering
\includegraphics[scale = 1.5]{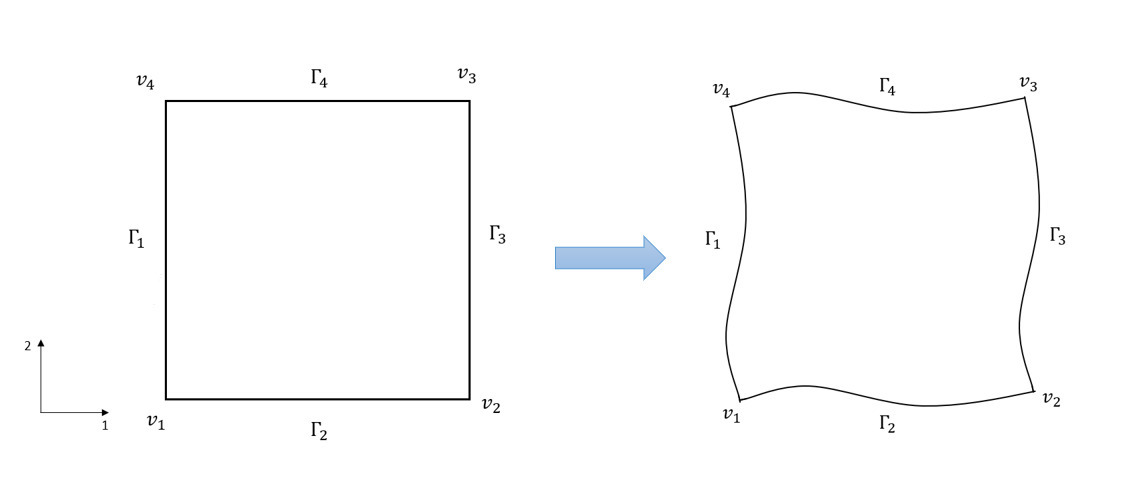}
\caption{Periodically deformed unit cell under general loading.}
\label{Fig.3}
\end{figure}

     The prediction of homogenised macroscopic properties of the composite materials for a given microstructure are central for designing the optimal composite materials. Although various theoretical models are available for prediction of effective elastic properties, they intrinsically take various assumptions, as given in \cite{b1,b2}. %(some references will be better here.). 
    Thus these theoretical models fail to account for true dependence of elastic properties on the spatial configuration of the matrix and the fiber constituting the microstructure.  Hence, in general, we rely on various computational approaches to evaluate the homogeneous elastic properties of composites for the given arbitrary microstructure. In the most popular computational approach, a finite element analysis is performed on a representative volume element (RVE) with periodic boundary conditions \cite{r1,r2}. For a binary composite material, a typical RVE is composed of fibre and matrix elements arranged spatially, as shown in Fig. \ref{Fig.1}.
    %A macroscopic composite material can be modelled as periodic stacking %of large number of similar microstructures, also known as the %representative volume element (RVE). For a binary composite material, %an RVE is composed of fibre and matrix elements arranged spatially, %as shown in Fig. \ref{Fig.1}.
 In this work, the composite material is modelled as a 2D plane strain system and both the fiber and matrix elements are assumed to be linear elastic and isotropic. Further, the fiber geometry is considered to be square in nature (see Fig.~\ref{Fig.1}). The following effective elastic constants for a given micro-structure of the composite material are evaluated.
    \begin{equation}
        \boldsymbol{E}=[E_{11}, E_{22}, G_{12}].
        \label{eq:1}
    \end{equation}
To determine these three elastic constants, the FEM calculations with three different quasi-static loadings are performed on the RVE. The considered loading cases are  longitudinal tension, transverse tension and pure shear as shown in Fig.\ref{Fig.2}. 
In all three cases, the periodic boundary conditions (PBCs) are imposed on the RVE (see Fig. \ref{Fig.3}). For example, longitudinal tension in `1' direction, the periodic boundary conditions can be given by:
\begin{align}
    u_{1,\Gamma 3}-u_{1,\Gamma 1} &= a\epsilon_{11},\label{eq:2} \\
     u_{2,\Gamma 4}-u_{2,\Gamma 2} &= 0,\label{eq:3} \\
    u_{1,\Gamma 4}-u_{1,v4} &= u_{1,\Gamma 2}-u_{1,v1},\label{eq:4} \\
    u_{2,\Gamma 1}-u_{2,v1} &= u_{2,\Gamma 3}-u_{2,v2},\label{eq:5}
\end{align}
where \(a\) is the edge length of the unit cell, (\(u_1,u_2\)) are displacements in `1' and `2' directions, \(\epsilon_{11}\) is the applied displacement constraint, \(\Gamma_i\) are boundary edges and, \(v_i\) are vertices. The displacements are shown in a deformed RVE in Fig.\ref{Fig.2}, and as shown, the boundaries \(\Gamma_1\) and \(\Gamma_3\) form a set which deforms together, and similarly, the boundaries \(\Gamma_2\) and \(\Gamma_4\) deform together due to applied PBCs. More details about the application of PBCs can be found in \cite{r2}.
For computational analysis of the RVE, the finite element analysis with 4-node plane strain elements have been performed. To evaluate effective material properties, the strain energy equivalence between the actual heterogeneous RVE and the effective homogeneous composite material has been used. The macro level stresses and strains in a 2D composite of area \(A\) are given by,
   \begin{align}
   \bar{\sigma}_{ij}  & = \frac{1}{A}\int_{A}^{}\sigma_{ij}(x,y)dA,\label{eq:6} \\
   \bar{\epsilon}_{ij}  & = \frac{1}{A}\int_{A}^{}\epsilon_{ij}(x,y)dA.\label{eq:7}
   \end{align} 
   Here, \(\sigma_{ij}(x,y)\) and \(\epsilon_{ij}(x,y)\) denote the stresses and strains in the composite material, respectively.
   The total strain energy \(U\) and \(U^*\) of the effective homogeneous composite material and the heterogeneous RVE over the area \(A\) can be given by
   \begin{align}
   U & = \frac{1}{2}\bar{\sigma}_{ij}\bar{\epsilon}_{ij}A,\label{eq:8} \\
   U^* & = \frac{1}{2}\int_{A}^{}\sigma_{ij}\epsilon_{ij}dA.\label{eq:9}
   \end{align}
   %\begin{align}
   %U^* & = \frac{1}{2}\int_{V}^{}\sigma_{ij}\epsilon_{ij}dV \\
   %U^* & = \frac{1}{2}\int_{V}^{}\sigma_{ij}(\epsilon_{ij}-\bar{\epsilon}_{ij}+\bar{\epsilon}_{ij})dV \\
   %& = \frac{1}{2}\int_{V}^{}\sigma_{ij}(\epsilon_{ij}-\bar{\epsilon}_{ij%})dV + \frac{1}{2}\bar{\epsilon}_{ij}\int_{V}^{}\sigma_{ij}dV \\
   %& = \frac{1}{2}\int_{V}^{}\sigma_{ij}\Big(\frac{\partial \bar{u}_i}{\partial x_j} - \frac{\partial u_i}{\partial x_j}\Big)dV + \frac{1}{2}\bar{\epsilon}_{ij}\int_{V}^{}\sigma_{ij}dV
   %\end{align} \\
  
   %Now, from (3) and (7), we have
   %\begin{align}
%       U^* - U &= \frac{1}{2}\int_{V}^{}\sigma_{ij}\Big(\frac{\partial %\bar{u}_i}{\partial x_j} - \frac{\partial u_i}{\partial x_j}\Big)dV \\
 %      U^* - U &= \frac{1}{2}\int_{V}^{}\frac{\partial[\sigma_{ij}(\bar{u}_i - u_i)]}{\partial x_j}dV
 %  \end{align}
  %  Result (9) holds because of the equilibrium relation
  % \begin{align}
  %    \frac{\partial \sigma{ij}}{\partial x_j} = 0
  % \end{align}
  % Using Gauss theorem, we get
  % \begin{align}
  %     U^* - U = \frac{1}{2}\int_{V}^{}\sigma_{ij}(\bar{u}_i - %u_i)n_{j}dS
  % \end{align}
  Using Gauss theorem and periodicity of boundary conditions,  we can show \(U^* = U\)\cite{r1}. Hence the average stresses and strains result in the equivalence in strain energies of the RVE and the overall homogeneous material. The ratio of these averaged stresses and strains along a particular direction will give us the effective elastic modulus in that direction. After describing the method to find the effective elastic properties through FEM, in the next section we discuss about the neural network based surrogate model to predict the effective elastic properties.
   %On the surface \(S\), \(u = \bar{u}\), and hence, \(U^* = U\). 
   
   %The stiff fiber has an Young's modulus of value E = 1 GPa and the soft matrix has E = 0.1 GPa. The Poisson's ratio \(\nu\) of both are taken to be equal to 0.33. 

\subsection{Convolutional Neural Network model} \label{CNN}
\begin{figure}[H]
\centering
\includegraphics[scale = 1.5]{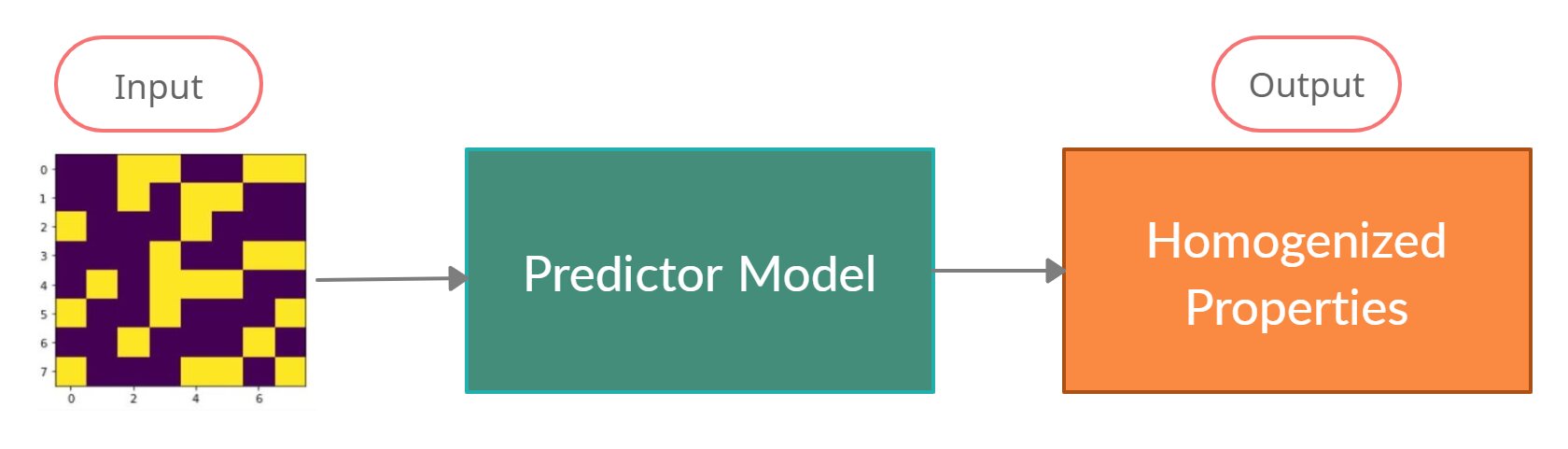}
\caption{Block diagram for the elastic properties predictor model.}
\label{Fig.4}
\end{figure}

\begin{figure}[H]
\centering
\includegraphics[scale = .75]{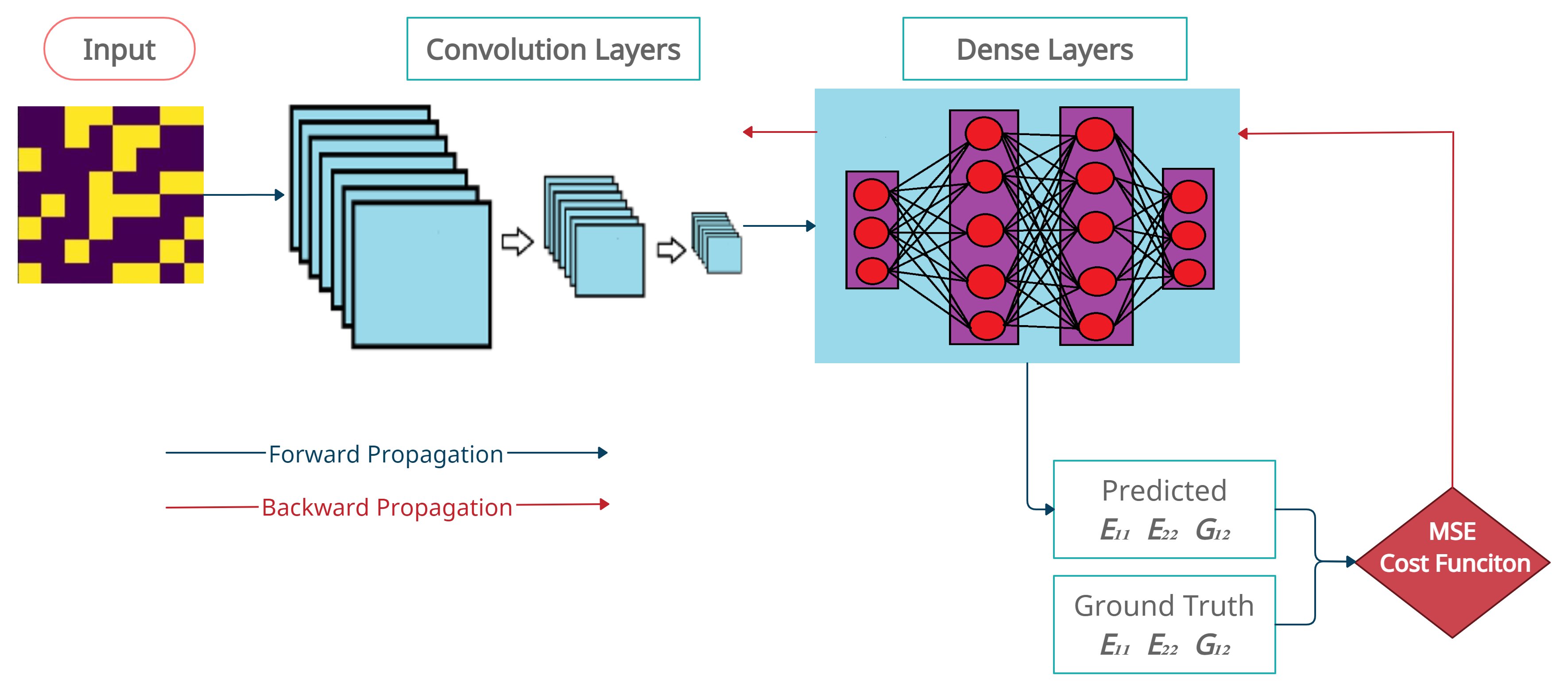}
\caption{Architecture of the deep neural network for the elastic property predictor model .}
\label{Fig.5}
\end{figure}

In the previous section, we observed that composite materials can be modelled using RVEs which map to modulus properties \(E_{11}\), \(E_{22}\) and \(G_{12}\). The values corresponding to these properties are calculated using FEM. However, FEM is a computationally expensive and time-consuming tool for calculating modulus properties. Hence there is need for a faster alternative for carrying out the analysis. Convolutional neural networks are a class of machine learning algorithms that can extract features from images and map them to suitable output. Due to their effectiveness, CNNs are used widely in domains such as facial recognition for biometric verification, analysing documents with tremendous accuracy, object detection, Natural Language Processing and many more cutting-edge technologies. Composite RVEs with square fibers can be represented in the form of checkerboard square images. The pixel values of the checkerboard image are assigned the value zero or one for representing the presence of fiber and matrix respectively. As the composite RVEs can be represented in the form of an image, CNNs can be used for predicting the modulus properties of composite RVEs. The CNN model takes composite RVE image as input and gives the modulus properties \(E_{11}\), \(E_{22}\) and \(G_{12}\) as output, as shown in Fig.\ref{Fig.4}.
The detailed architecture of the deep neural network for the elastic property predictor model is depicted in Fig.~\ref{Fig.5}. The input microstructure is fed to a series of convolutional layers which are further connected to a few dense layers. The last dense layer has three nodes and  the output of these nodes provide the prediction of the elastic modulus (\(E_{11}\), \(E_{22}\) and, \(G_{12}\)). The weights of the convolutional neural networks layers are obtained by minimising a cost function. Here, since the  objective is to minimize the difference between ground truth (FEM data) and CNN predicted values, we have employed the Mean-Squared-Error (MSE) as our cost function.
%Convolutional neural networks work on a gradient based optimization algorithms which performs repeated iterations to reduce the difference between ground truth and CNN predicted values, which is defined by a cost function.
\begin{align}
    MSE = \frac{1}{n}\sum_{i = 1}^{n}(y_i - \hat{y}_i)^2,\label{eq:10}
\end{align}
where \(y_i\) is model output from the last dense layer and \(\hat{y}_i\) is the ground truth value obtained from FEM. For testing the accuracy of the model, we use Mean Absolute Percentage Error (MAPE) which is given by
\begin{align}
    MAPE = \frac{1}{n}\sum_{i=1}^{n}\mid{\frac{y_i - \hat{y}_i}{y_i}}\mid{}.\label{eq:11}
\end{align}
The CNN model is trained with the ground truth data obtained from FEM analysis with the help of gradient based optimization scheme.  After the training process, the CNN model is used for carrying out the property prediction of the Composite RVEs.
\subsection{Data augmentation} \label{DATA}
We have explained in the last subsection that CNN can be used for predicting the effective elastic properties of composite materials. The performance of the CNN model is measured by the difference between CNN predictions and ground truth FEM data. Their performance is dependant on a number of factors, and dataset is one of the most significant one~\cite{p13}. Increasing the dataset size can significantly enhance the prediction accuracy of the machine learning model \cite{p11}. Generating training data of sufficient size from FEM can be computationally expensive and time consuming and may require the use of HPCs (High Performance Computers). To overcome this limitation, we propose a novel data augmentation strategy for increasing the dataset size and thus  significantly bringing down the computational resources required during the training phase. The overall data augmentation procedure is grouped into two categories: 
\begin{itemize}
    \item Flip, Rotate and Transpose
    \item Periodic augmentation
\end{itemize}
The first scheme employs elementary operations which are commonly used in data augmentation methods used in image processing. While in the second scheme, we are exploiting the periodicity of composite materials to increase the size of the dataset. Note, that since the data augmentation rely on only the periodicity of the micro-structure, the strategy can be used in any application involving the periodic micro-structures. These procedures are explained in detail in below subsections.
\subsubsection{ Flip, Rotate and Transpose}
\begin{figure}[H]
\centering
\includegraphics[scale = 2.2]{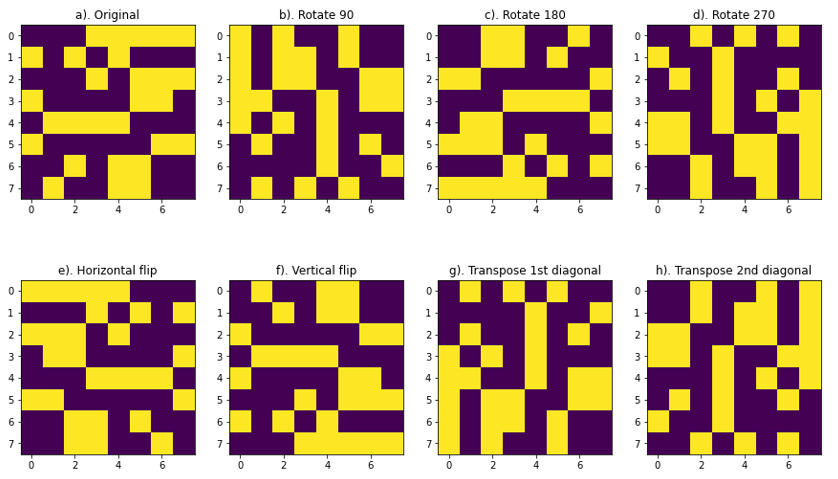}
\caption{Different types of geometrical augmentations.}
\label{Fig.6}
\end{figure}

In this section, we demonstrate how performing the operations of rotating, flipping and taking transpose of the RVE image can increase our dataset size. A sample \(8\times 8\) composite design having \(E_{11} = a\) GPa and \(E_{22} = b\) GPa is represented in \ref{Fig.6}(a). When we rotate the original image in Fig.\ref{Fig.6}a by 90°, we get the image shown in Fig.\ref{Fig.6}b. The modulus properties of this image will get interchanged, i.e., \(E_{11} = b\) GPa and \(E_{22} = a\) GPa  for Fig.\ref{Fig.6}b. When we rotate the original image by 180 degrees (Fig.\ref{Fig.6}c), the modulus properties will remain same as the original image. On rotating by 270 degrees (Figure 5d), \(E_{11}\) and \(E_{22}\) will get interchanged. Similarly, when we flip the image about a horizontal axis (Fig.\ref{Fig.6}e) or vertical axis (Fig.\ref{Fig.6}f). The modulus properties will remain same as original. Lastly, when we take the transpose of the image about the first diagonal (Fig.\ref{Fig.6}g) or second diagonal (Fig.\ref{Fig.6}h), the properties \(E_{11}\) and \(E_{22}\) will get interchanged.
To summarize: when we rotate the image by 180 degrees or flip it horizontally or vertically, the properties will remain same as original. When we rotate by 90 degrees or 270 degrees or take the transpose about any of the diagonals, the properties will get interchanged. These newly augmented RVE's have been validated using FEM analysis, hence these are valid composite designs which can be added to our original dataset. This procedure increases the size of base dataset by a factor of eight.
\subsubsection{ Periodic augmentation}
\begin{figure}[H]
\centering
\includegraphics[scale = 2.2]{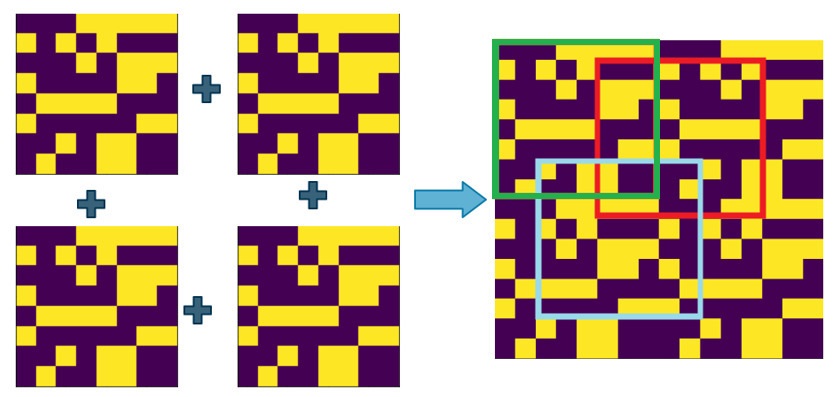}
\caption{Data augmentation by exploiting the periodic nature of the RVEs. Different coloured \(8\times 8\)  RVEs  shown on the right side have the identical elastic properties.}
\label{Fig.7}
\end{figure}

In addition to previously discussed operations, the periodic stacking of RVEs also offers a technique for further augmentation of training dataset size. In Fig.~\ref{Fig.3}, the extended RVE formed by stacking four designs of Fig.\ref{Fig.1}a alongside each other is shown. The resultant RVE formed by stacking four \(8\times 8\) base RVEs together results in a final \(16\times 16\) image shown on the right side of Fig.\ref{Fig.7}. In the \(16\times 16\) RVE, three squares of green, red and blue boundaries are depicted with each of the squares enclosing an \(8\times 8\) image. The green square represents the original image from which we constructed the extended design. The red and blue squares represent some randomly drawn \(8\times 8\) squares on the extended design. It turns out, if we use either blue or red square instead of green square (which represents the original base RVE), we will get the same extended design as the one we constructed using green square. As the modulus properties of the extended \(16\times 16\) RVE is same as the \(8\times 8\) green RVE, the modulus properties of red and blue RVEs are also similar to modulus properties of green RVE as they all construct the same extended RVE. The \(8\times 8\) RVEs enclosed by red and blue boundaries are also valid RVEs that can be added to our base dataset.  The number of different \(8\times 8\) squares that we can cut out from a \(16\times 16\) image is equal to 64, which is simply square of the side of the base RVE. In general, for a base RVE of side dimension = \(N\), number of different \(N\) sided RVEs that we can get having same modulus properties is equal to \(N^2\). This procedure increases the size of base dataset by a factor of \(N^2\), where \(N\) is the side of the square image.
Both these augmentation procedures can be implemented simultaneously in various cases which involve the periodicity in their structures. Hence the resultant multiplying factor of our dataset by combining both augmentation procedures on a RVE of side length \(N\) is \(8N^2\). All the additionally generated RVE's have been validated using FEM analysis. Although there is a chance of some of these new RVEs being identical depending on the distribution of elements in the RVE, the chance or percentage of such happenings is minimal and can be ignored.\\
For an \(8\times 8\) RVE, the multiplying factor = \(8 \times (8)^2 = 512\).\\
For a \(16\times 16\) RVE the multiplication factor = \(8 \times (16)^2 = 2048\).

Thus, we can see that data augmentation technique can increase the data size significantly without almost no additional computational effort.
\subsection{ Generator model} \label{GEN}
As shown in the previous section, the proposed predictor model can be used as a surrogate model to FEM for accurate prediction of the mictrostructural properties. Hence, with the predictor model, we are able to predict the elastic properties precisely if the microstructure is known. However, one of the crucial aspects of composite design is to generate microstructures for some desired properties under some physical constraints. Brute force methods of generating microstructures are very tedious, since even an \(8\times 8\) binary RVE with 50\% fibre volume fraction can be designed in \(^{64}P_{32} \approx 10^{51}\), which can take decades using modern computers. Hence there is an immediate need for a faster and lightweight model for design and discovery of new microstructures for a given objective.

For the purpose of designing microstructures of desired properties,  we present a neural-network based  predictive-generative model. The developed predictive-generative model is a general purpose neural network framework and can generate the microstructure design by optimizing any objective function in terms of its elastic and geometric properties.

\begin{figure}[H]
\centering
\includegraphics[scale = 1.7]{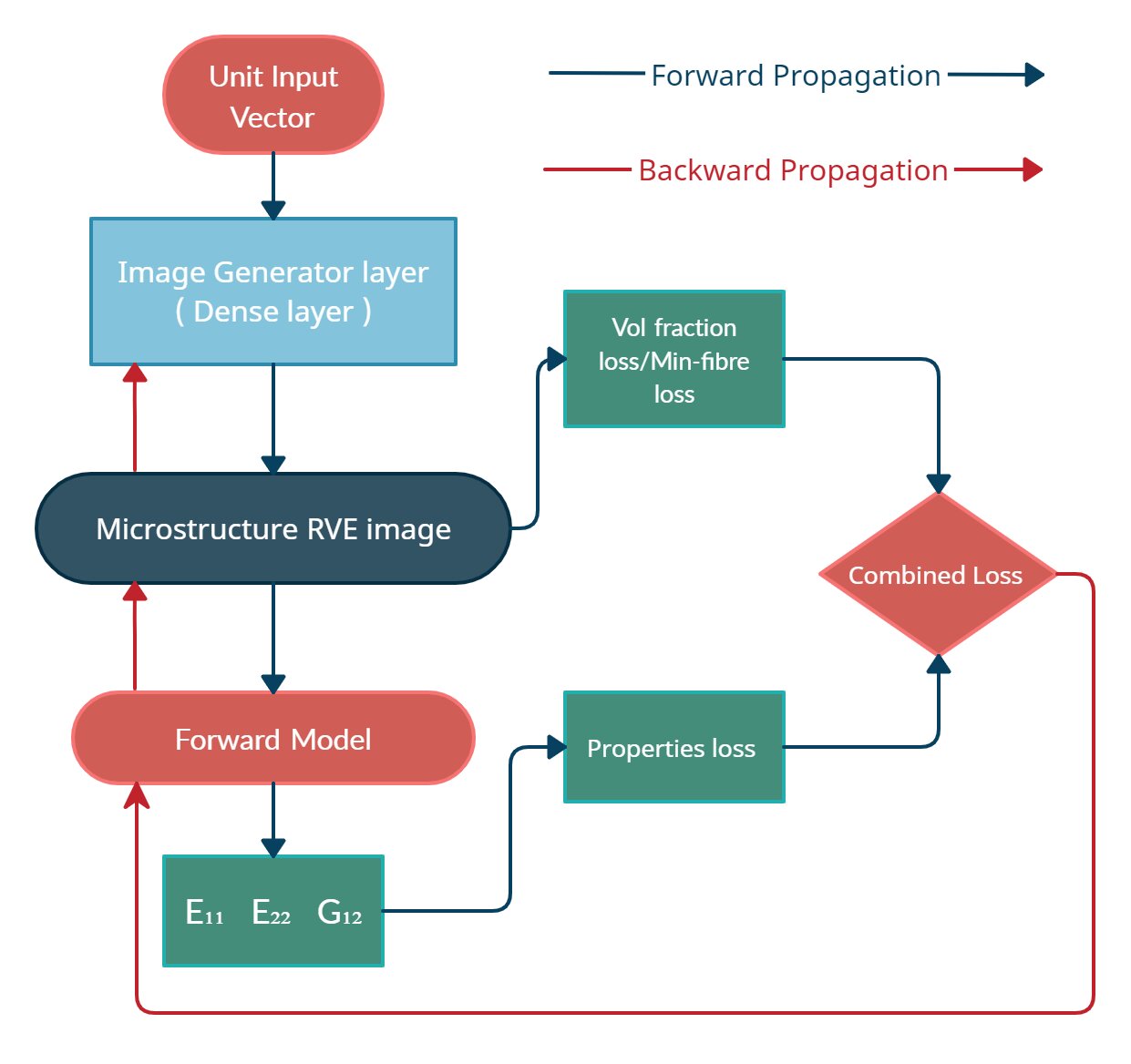}
\caption{Generative model: Schematic of the neural network based framework for the generation of the optimal composite microstructure.}
\label{Fig.8}
\end{figure}

The proposed framework of the generator model is shown in the Fig.~\ref{Fig.8}. As shown in the schematic, the generator model consist of a single dense layer, which is used to generate the microstructure of the RVE. The dense  is activated with the sigmoid activation function. The use of the sigmoid function ensures that the pixel value in the microstructure remains between zero and one. Further, the output from the dense layer is reshaped to obtain the 2-D microstructure image. This dense layer is connected to a unit input vector of 100 dense layer nodes. The microstructure generated from the generator layer is then fed to the predictor model, which is already trained and thus for any given input, predicts the elastic properties. Further, the desired objective function can be defined in terms of the predicted elastic properties. 
To obtain the optimized microstructure, the overall objective function is minimized through the gradient based optimization scheme. The gradients can be calculated by the standard back propagation scheme.  Note, since the weights and bias of the predictor model are pre-trained using the algorithm presented in section 2.2, the only trainable parameters in this model are the weights of the generator layer.
The loss function of the predictive-generative model considered in the manuscript can be broken in to three terms  \(L_p,L_v,L_u\) , which deal with output elastic moduli, fibre volume fraction and constraining the element pixel values to one and zero respectively. 
\begin{align}
       L = w_{1}L_p+ w_{2}L_v + w_{3}L_u.
       \label{eq:12}
\end{align}
Here, the first term \(L_p\) is defined in terms of the elastic properties, and used to ensure the desired elastic modulus response of the generated microstructure.  For example, to obtain the microstructure with the specified property (\(E_{11}^*,E_{22}^*)\),  $L_p$ can be taken as:
\begin{align}
    L_p = [(E_{11}-E_{11}^*)^2+(E_{22}-E_{22}^*)^2].\label{eq:13}
\end{align}
The second term, \(L_v\) is used to satisfy the constrain on volume fraction, for example specifying the maximum number of the fibers that are allowed in the micro-structure:
\begin{align}
       L_v = \Big(\sum_{i}{}[1-\rho_i] - \nu\Big)^2.\label{eq:14}
\end{align}
The last term, \(L_u\) is provided to ensure that the obtained micro-structure is binary in nature. The specific expression of this function, which ensures the pixel values to be either zero or one is given below.
\begin{align}
    L_u = \sum_{i} (\rho_i(\rho_i-1))^2.\label{eq:15}
\end{align}
The weights \(w_{j}\) can be chosen based on the importance of the each loss function term that the designer wants to provide. More details about the specific objective functions will be provided in the results and discussion section.
\section{ Results and Discussion} \label{RESULT}
In the previous sections, a predictive-generative model based on the deep neural networks is presented towards tailoring the composite micro-structure for desirable elastic properties or vice-versa. In this section, we present  various numerical examples to demonstrate the efficacy of the proposed approach. In the first example, we demonstrate the ability of the  proposed data augmentation technique in enhancing the performance of the predictive model. Further, we present examples to demonstrate the capacity of generative model to solve the inverse homogenisation problem for the desired elastic response. In addition, we also show that how different constraints can be incorporated in the generator model while designing the microstructures.

 In the numerical examples presented, a composite material consisting of the isotropic and elastic fiber is considered. The elastic modulus for the matrix and the fiber is taken as 0.1 GPa and 1 GPa respectively, while the Poission's ratio for both is taken as 0.33. The size of the square RVE considered is 20 \(\mu\)m. The RVE is being discretized into the grid of \(8\times 8\). In each grid cell, either the fiber or matrix can be present (see Fig.~\ref{Fig.4}).

\subsection{ FEM Analysis}
 For generation of the training data for the predictive model, FEM analysis using $4$-noded plane strain quadrilateral CPE4 elements is performed. The FEM simulations provide the value of the elastic constants for the given RVE. To ensure that our results are independent of mesh size, a mesh convergence study has been carried out. The data corresponding to the \(E_{11}\), is shown in Table~\ref{table:1}.
\begin{table}[H]
\centering
\begin{tabular}{c|c}
     \toprule
     Element number & \(E_{11} (GPa)\) \\
     \hline
     1600 & 0.32154  \\
     \hline
     6400 & 0.32137 \\
     \hline
     40000 & 0.32135 \\
     \hline
\end{tabular}
\caption{FEM mesh convergence for a typical microstructure.}
\label{table:1}
\end{table}
From Table~\ref{table:1}, it is seen that for an increase in number of elements from 6400 to 40000, \(E_{11}\) changes only by 0.006\%, hence we select our global mesh size to be 0.25 \(\mu m\), which corresponds to 6400 elements. The data generated using FEA will now be used to train the the predictor model, as discussed in the next section.  
% {\bf Fig. 9 is labelled twice.}
\subsection{ Predictor model}
\begin{figure}[H]
\centering
\includegraphics[scale = 1.4
]{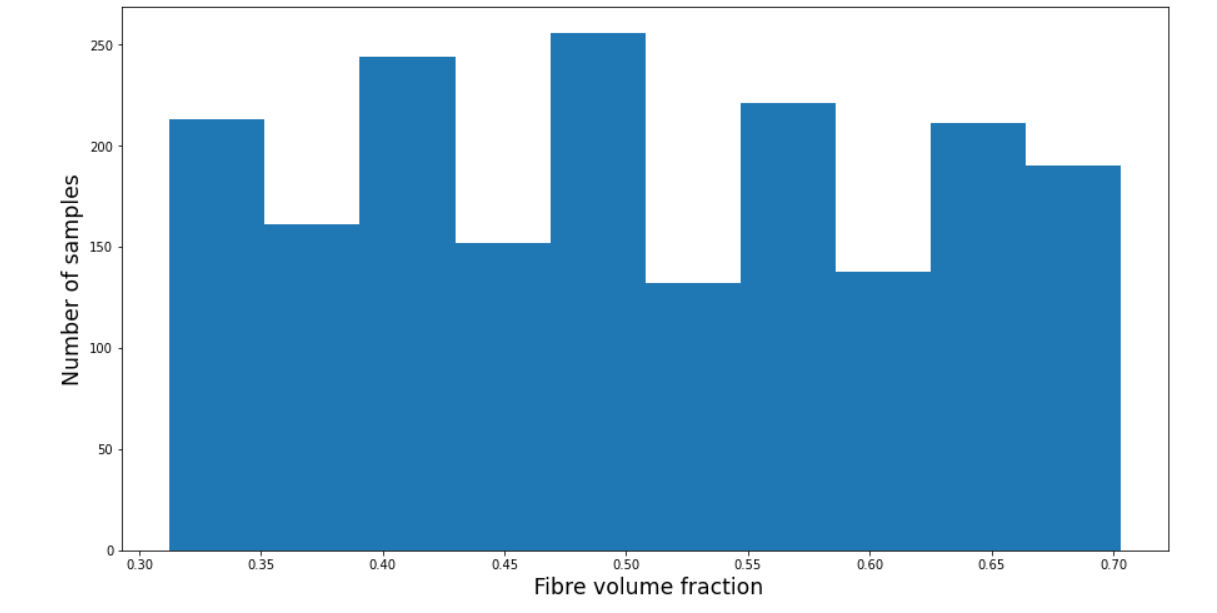}
\caption{Distribution of fibre volume fraction in the test data.}
\label{Fig.9}
\end{figure}

\begin{figure}[H]
\centering
\includegraphics[scale = 1.55]{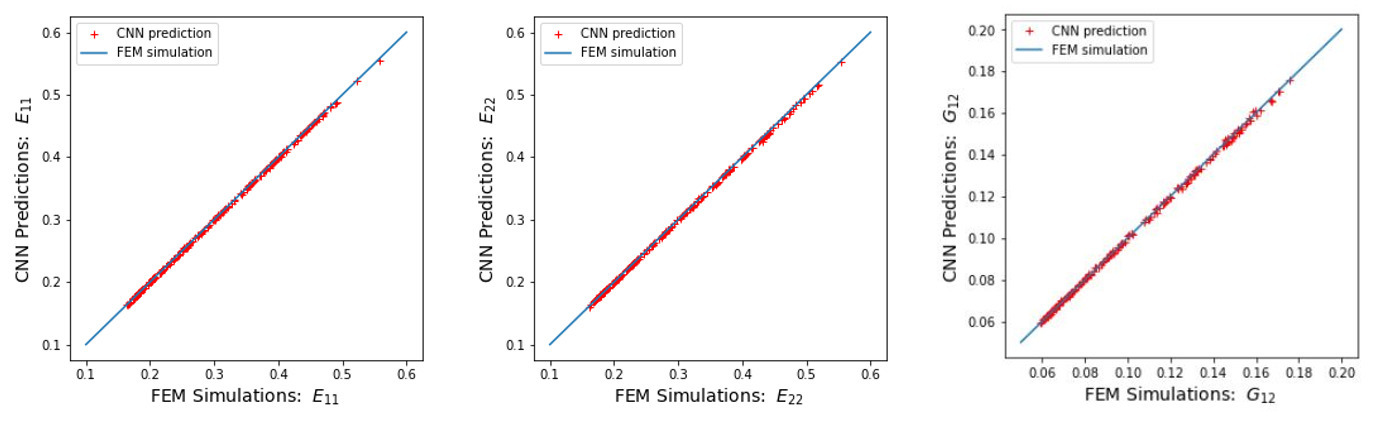}
\caption{Comparison of neural network based predictor model with the FEA for 200 test samples.}
\label{Fig.10}
\end{figure}

As discussed in the the theory section, the predictor model is employed to predict the elastic properties of composite materials. For training the CNN based predictor model, we generated a ground truth dataset of size 9000 from FEM analysis taking a variable fibre volume fraction between 0.3 and 0.7. The data augmentation procedure detailed in section 2.3 is applied to this base data, and after augmentation,  a final dataset of around 4.6 Million is obtained.  
The programming framework that we have used to deploy the CNN model is TensorFlow using the Keras library. The optimizer used is ADAM (adaptive moment estimation). Further details of the model architecture for the predictor model are listed in Table~\ref{table:2}.
\begin{table}[H]
\centering
\begin{tabular}{c|c}
     \toprule
     Hyperparameter & Value \\
     \hline
     Number of Convolution Layers & 5, with SAME padding \\
     \hline
     Number of Dense Layers & 4 \\
     \hline
     Number of Nodes & (256,256,128,128) \\
     \hline
     Learning Rate & 0.001 \\
     \hline
     Activation Function & ReLu (intermediate), linear (output) \\
     \hline
     Minibatch size & 256 \\
     \hline
\end{tabular}
\caption{Model hyperparameters for property predictor model.}
\label{table:2}
\end{table}
We trained the CNN model for two cases, one with the base dataset and another with the augmented dataset. A test dataset consisting of 2000 samples is generated using FEM and model performance for both cases is tested on this new data. Testing the model on this test data ensures that the model has not been trained on samples of this test data set or any of its corresponding augmented samples. In Fig.~\ref{Fig.9}, the distribution of the various fibre volume fractions in the test dataset is shown. While the model trained on the base dataset, {\it i.e.} before the augmentation procedure has an error of 2.8\%,  the model trained on augmented dataset, {\it i.e.} after augmentation shows the error of 0.40\%. This comparison validates the effectiveness of the developed predictor model.
%The significant reduction in test error due to data augmentation demonstrates the utility of the proposed procedure.
 Table~ \ref{table:3} shows various performance measures of the CNN model trained before and after augmentation procedure. The results depict a considerable improvement in the model performance after the augmentation procedure.
\begin{table}[H]
\centering
\begin{tabular}{c|c|c}
     \toprule
     Parameter & Before augmentation & After augmentation \\
     \hline
     Train Dataset size & 9000 & 4.6 million \\
     \hline
     Training error & 0.33\% & 0.39\% \\
     \hline
     Test error & 2.8\% & 0.40\% \\
     \hline
     \(R^2\) score & 0.9819 & 0.9997 \\
     \hline
     Samples with error\( >\) 2\% & 21\% & 0.0001\%
     
\end{tabular}
\caption{Effect of data augmentation on the performance of the property predictor model.}
\label{table:3}
\end{table}
The obtained predictions with model trained on the augmented data are shown in Fig.\ref{Fig.10} %{\bf wrong figure reference i think.} 
and Table~\ref{table:4}. As can be seen from Fig.\ref{Fig.10}, the prediction made by the CNN model are very close to FEM.  In Table~\ref{table:4}, we can observe the minimal difference between the model predictions and FEM values for a few microstructures. 

\begin{table}[H]
\centering
\begin{tabular}{c|c|c}
     \toprule
     FEM output \([E_{11},E_{22},G_{12}]\) & Microstructure & Predicted output \([E_{11},E_{22},G_{12}]\) \\
     \hline
     [0.1813, 0.1917, 0.0691] GPa & 
     \includegraphics[width=0.17\textwidth, height=28mm]{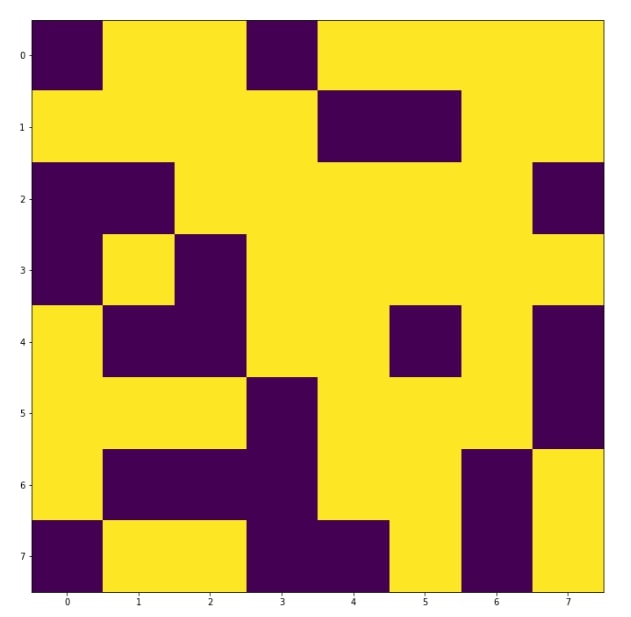} &
     [0.1823, 0.1930, 0.0686] GPa \\
     \hline
     [0.2546,0.2921,0.0938] GPa &
     \includegraphics[width=0.17\textwidth, height=28mm]{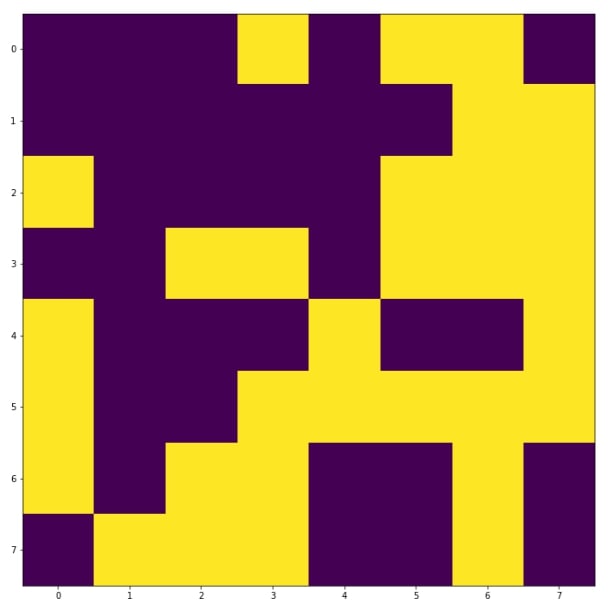} &
     [0.2562,0.2924,0.0938] GPa \\
     \hline
     [0.2410, 0.2846, 0.0992] GPa &
     \includegraphics[width=0.17\textwidth, height=28mm]{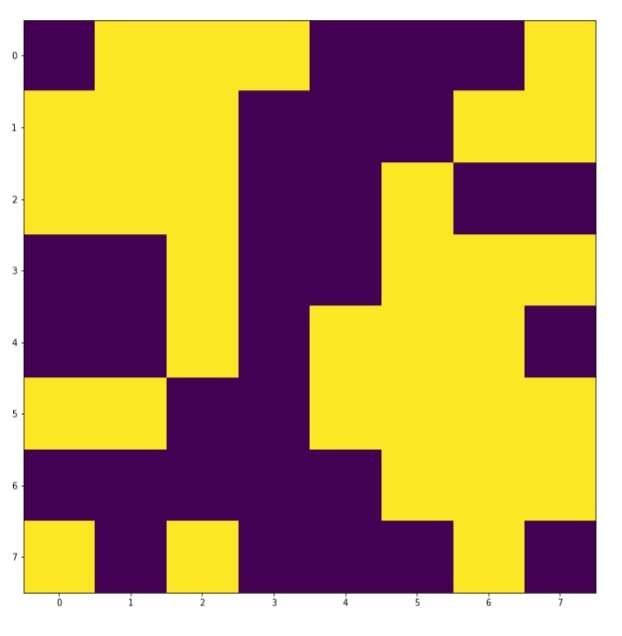} &
     [0.2410, 0.2840, 0.0993] GPa \\
     \hline
\end{tabular}
\caption{Comparison of FEM and CNN predicted output for some microstructures.}
\label{table:4}
\end{table}
%\begin{figure}[H]
%\centering
%\includegraphics[scale = 0.8]{Hyperparameters.jpeg}
%\caption{Model hyperparameters.}
%\end{figure}
% After the training procedure, new test data consisting of 2000 samples was generated using FEM and model performance was tested on this new data. Testing the model on this new test data ensures that the model had not been trained on samples of the new test data or any of its corresponding augmented samples. The error on this new data was less than 0.35\%, which validates the effectiveness of the developed predictor model. Table 4 shows various performance measures of the CNN model trained before and after augmentation procedure. The results depict a drastic improvement in the model performance after the augmentation procedure.
% \begin{table}[H]
% \centering
% \begin{tabular}{c|c|c}
%      \toprule
%      Parameter & Before augmentation & After augmentation \\
%      \hline
%      Dataset size & 10500 & 5 million \\
%      \hline
%      Training error & 0.33\% & 0.39\% \\
%      \hline
%      Test error & 2.8\% & 0.40\% \\
%      \hline
%      \(R^2\) score & 0.9819 & 0.9997 \\
%      \hline
%      Samples with error\( >\) 2\% & 21\% & 0.0001\%
     
% \end{tabular}
% \caption{Effect of augmentation.}
% \label{table:4}
% \end{table}

%\begin{figure}[H]
%\centering
%\includegraphics[scale = 0.8]{augmentation.jpeg}
%\caption{Effect of augmentation.}
%\end{figure}
\subsection{ Inverse model}
 After establishing the accuracy and predicting capabilities of predictor model, it is combined with an inverse generator model to design the composite microstructures. In this section, we will discuss the efficacy of the generator model for various settings of increasing complexity. We begin with a simple case, for which the results can be verified by simple intuition itself. After that, we present the cases, where the generator model is used for designing microstructure for user defined elastic properties. At last, we try to minimize the fiber content in the microstructure subject to user defined elastic properties.
%Microstructures designed using these constraints can have tremendous applications in optimizing strength in either directions, weight and cost reduction and so on. 

As discussed previously, the constraints will be applied through loss functions which will be minimized by the neural network. The elastic modulus loss function \(L_p\) and the volume loss function \(L_v\) will be used in both of the above described settings as an optimizer or a constraint. The loss function \(L_u\) has been described in detail in the theory section.
\subsubsection{ Elastic Modulus optimizer}
 In this section, we start with the simplified setting, where the results obtained from the generator model can be verified intuitively. In this simplified setting for the inverse model, we take the following expression of the elastic modulus loss function \(L_p\) as a part of the total loss function \eqref{eq:12} 
\begin{equation}
    L_{p} = E_{mm}-E_{tt}.\label{eq:16}
\end{equation}  
Here, \(E_{mm}\) and \(E_{tt}\) denote elastic modulus in `$m$' and `$t$' directions respectively. Since, we are minimizing the total loss, it is expected to give microstructure designs  with the maximum elastic modulus in `\({t}\)' direction, while minimum modulus in  `\({m}\)' direction. We know intuitively that the micro-structure for this objective function is expected to align the fibers in the `\({t}\)' direction. Further, the fiber volume constraint is implemented by taking the $L_{\nu}$ in the total loss function \eqref{eq:12} is taken as:
\begin{equation}
       L_{\nu} = \Big(\sum{}{}\rho_i - \nu\Big)^2.\label{eq:17}
\end{equation}

\begin{figure}[H]
\centering
\includegraphics[scale = 0.9]{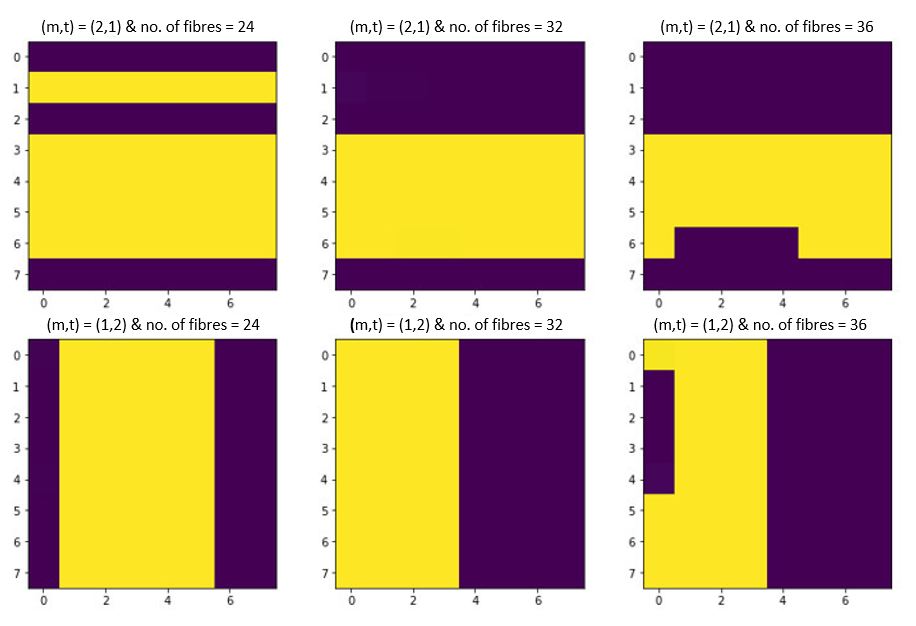}
\caption{A few optimal microstructures obtained from the generative model with elastic modulus optimized setting.}
\label{Fig.11}
\end{figure}

Here, \(\nu\) is the desired number of matrix elements and \(\rho_i\) are the image pixel values of the RVE denoting the the matrix present at a particular grid. The microstructure designs obtained by the minimizing the multi-objective expression are expected to give the microstructure aligned in the `$t$' direction and satisfying the fiber volume criterion. The designs generated for the different fiber volume fraction are shown in Fig.~\ref{Fig.11},. The top row consist of the microstructures predicted by maximizing the  elastic modulus along `$1$' and minimizing along `$2$'. While, the bottom row shows the predictions for the cases when the elastic modulus is maximized in the direction of `$2$' and minimized along `$1$'. The designs obtained are consistent with the intuition that the fibers will try to align themselves in the direction  `$t$'. Thus through this simple example, we are able to verify the implementation of the proposed generator model.
\subsubsection{ Desired Elastic properties}
In this subsection, we consider a setting of designing the micro-structure of the composite material for obtaining the specific elastic  properties. In this setting, to obtain the microstructure with the specified property (\(E_{11}^*,E_{22}^*)\), the $L_p$ in the total loss function \eqref{eq:12} is taken as:
\begin{align}
    L_p = [(E_{11}-E_{11}^*)^2+(E_{22}-E_{22}^*)^2].\label{eq:18}
\end{align}
In this case, since no volume constraint is considered the overall loss function consists of only the \(L_p\) and \(L_u\) and $w_2$ is taken as zero.  Some of the microstructures generated using our predictive-generative model for the specified elastic properties case are shown in Table~\ref{table:5}. The desired specified properties are listed in the left column of the table and the predicted microstructure are shown in the middle column. The right most column of the table shows the properties associated with the predicted microstructure. We can see from the table that the micro-structure obtained from the generator model indeed provide the desired elastic response. In Table~\ref{table:5}(a) and (d), equal elastic properties are specified in both the direction, while in Table~\ref{table:5}(b) and (c) the properties are asymmetric in nature. In these asymmetric properties case, the microstructure also tend to be asymmetric in nature and the fibers are aligned in the direction of the higher elastic modulus. The generated microstructures are also validated with FEM simulations and the output properties are found to be in very good agreement. We used the volume fraction from the predicted microstructures to evaluate the resultant properties using the Rule of Mixtures. For the first case, the theoretical predictions were (0.761,0.295) GPa, which is significantly different from the actual properties used in the model. This contrast is observed in the other cases too, for example (0.748, 0.283) GPa for case 2, (0.803,0.337) GPa for case 3 and (0.748,0.283) GPa for case 4. Hence, we can say that the model is able to accurately design microstructures with certain properties which would be quite difficult to discover using the existing theoretical models. The generator model has designed these microstructures with minimal computational effort (in personal computer less than half a minute). (Note that for the microstructures in Table \ref{table:5} and the subsequent figures, black represents fiber and white represents the matrix elements).  

\begin{table}[H]
\centering
\scalebox{0.9}{
\begin{tabular}{c|c|c}
     \toprule
     Given elastic modulus \([E_{11},E_{22}]\) & Predicted Microstructure & Output elastic modulus \([E_{11},E_{22}]\) \\
     \hline
     [0.4, 0.4] GPa & 
     \includegraphics[width=0.17\textwidth, height=28mm]{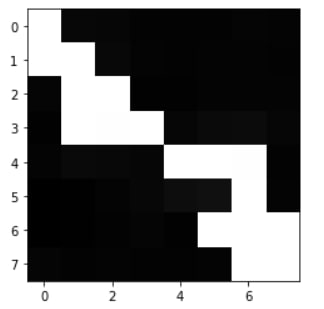} &
     [0.4011, 0.399] GPa \\
     \hline
     [0.5,0.4] GPa &
     \includegraphics[width=0.17\textwidth, height=28mm]{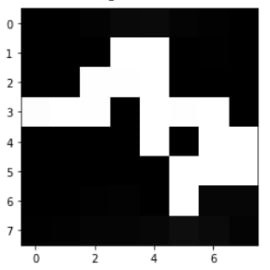} &
     [0.507,0.394] GPa \\
     \hline
     [0.6, 0.4] GPa &
     \includegraphics[width=0.17\textwidth, height=28mm]{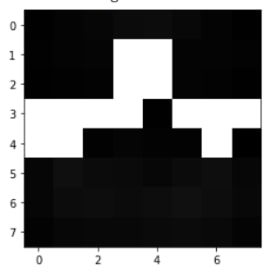} &
     [0.608, 0.401] GPa \\
     \hline
     [0.5, 0.5] GPa &
     \includegraphics[width=0.17\textwidth, height=28mm]{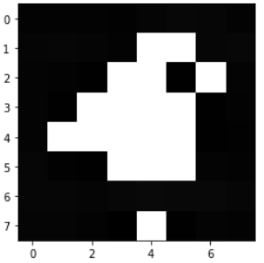} &
     [0.505, 0.505] GPa \\
     \hline
\end{tabular}}
\caption{A few predicted microstructures for given elastic properties without volume constraint.}
\label{table:5}
\end{table}

\begin{figure}[H]
\centering
\includegraphics[scale = 0.85]{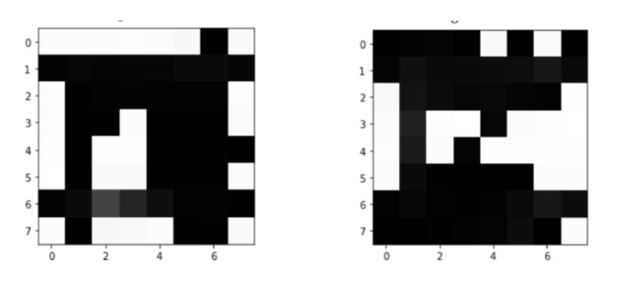}
\caption{Predicted microstructures with volume constraint for modulus a) (\(E_{11},E_{22}\)) = (0.4, 0.4) GPa, \(\nu_f\) = 62.5\%  b) (\(E_{11},E_{22}\)) = (0.5, 0.4) GPa, \(\nu_f\) = 67.2\%.}
\label{Fig.12}
\end{figure}

%\subsubsection{\large Designing microstructures of desired properties and specific volume fraction.}
Now, once we know that the model can accurately design microstructures for a given elastic properties, we extend the model's capabilities to design microstructures of desired properties and having a specific input fibre volume fraction. The specified fibre volume fraction will be used as a constraint along with the property loss function. In Fig.\ref{Fig.12},  microstructures are designed for (\(E_{11},E_{22}\)) = (0.4, 0.4) GPa and (\(E_{11},E_{22}\)) = (0.5, 0.4) GPa for a specified volume fraction of 62.5\% and 67.2\% respectively. We can see that with the volume constraint, the obtained microstructure for the same elastic properties, is different from the one obtained in table~\ref{table:5}.

\subsubsection{ Fibre volume fraction optimizer}
 In this subsection, we use the generator model to obtain the micro-structure of the composite materials of the desired elastic response with the least possible content of the fiber material. The designing of matrix-fiber configurations for a specified elastic modulus with minimal fiber material content can be useful in enhancing the multifunctional capabilities of the composite materials while ensuring sufficient stiffness. %({\bf a bit lost in this line.}).

To predict the microstructures with minimum fibre content, the following expression is used in the total loss function:
\begin{align}
       L_v = \sum_{i=1}^{n}{(1-\rho_i)^2},\label{eq:19}
\end{align}
Hence, minimization of the objective function ensures that fiber volume fraction is minimized, while satisfying the elastic property loss function.   
\begin{figure}[H]
\centering
\includegraphics[scale = 1.6]{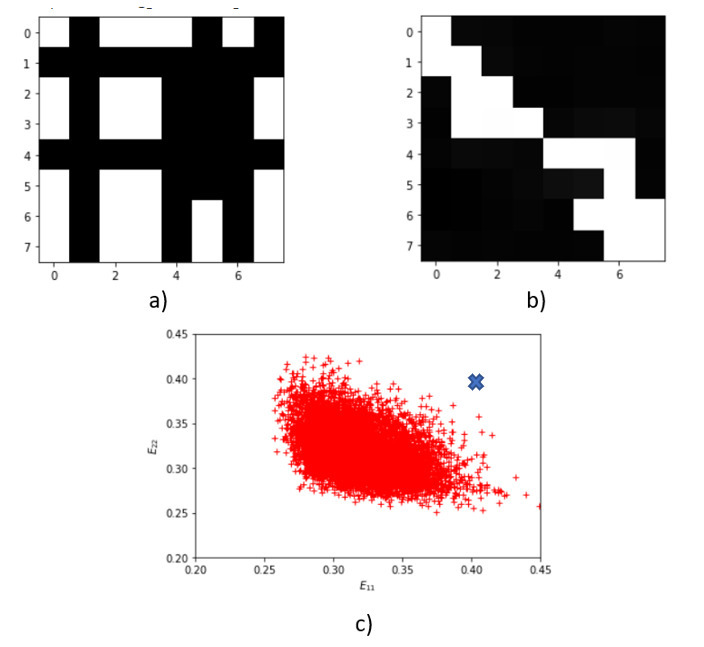}
\caption{Predicted microstructures and scatter plots for elastic modulus of (0.4, 0.4) GPa, a) Using volume fraction optimizer, b) Without using any volume loss function, and c) Scatter plot obtained at minimum volume fraction.}
\label{Fig.13}
\end{figure}
\begin{figure}[H]
\centering
\includegraphics[scale = 1.5]{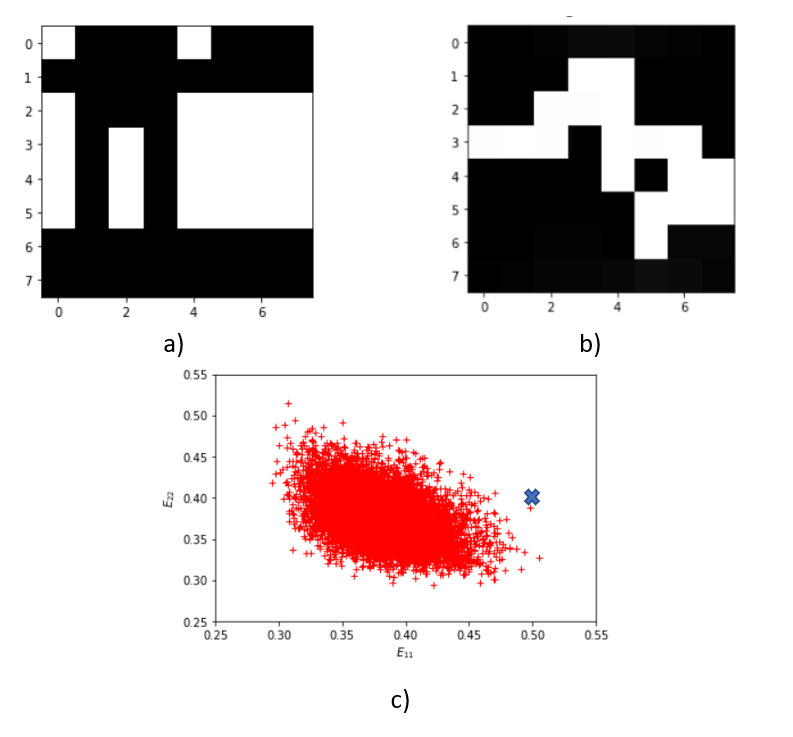}
\caption{Predicted microstructures and scatter plots for elastic modulus of (0.5, 0.4) GPa, a) Using volume fraction optimizer, b) Without using any volume loss function, and c) Scatter plot obtained at minimum volume fraction.}
\label{Fig.14}
\end{figure}
The microstructure obtained by generator model for the minimum fiber material for an input modulus of (0.4, 0.4) GPa is shown in Fig.\ref{Fig.13}a. The fiber volume fraction of the obtained design is 57.8\%, while the volume fraction of the design (Fig.\ref{Fig.13}b) obtained without using any volume fraction loss function is 73.4\%. Thus, it can be observed that the resulting volume fraction using the optimizer is much lower than the microstructure obtained without it. To further demonstrate the efficacy of the proposed approach,the scatter plot for the fiber volume fraction of 57.8\% is shown in Fig.\ref{Fig.13}c. This scatter plot is obtained using the randomly generated designs and using the predictor model. The design in Fig.~\ref{Fig.13}a proposed by the generator for the minimum fiber case, is marked in the scatter plot with the cross sign. %({\bf can not see the cross.}). 
It can be seen that this design is one of the most extreme cases at this volume fraction. This further verifies that the design obtained indeed is one with the minimum volume fraction.

In Fig.\ref{Fig.14}, another example of minimum fibre volume fraction is shown, for input properties of (0.5,0.4) GPa. In this case, the optimizer discovers a configuration at 60.9\% fiber fraction, whereas the model creates a geometry at 71.85\% without using the fiber volume optimizer. The resulting scatter plot in Fig.\ref{Fig.14}c also highlights that the resulting microstructure is an extreme case. This proves that the inverse model can efficiently generate borderline cases at the extreme ends of the distribution by creating microstructures which are difficult to conceptualize otherwise.

{%\\[0.5cm]\large Fig.\ref{Fig.12}a shows the the scatter plot of property predictions at a given fibre volume fraction of 62.5\% and the microstructure obtained. It can be easily observed that to obtain a desired modulus of (0.4,0.4) GPa, this configuration is not optimal in terms of fibre volume fraction since the given properties lie almost at the centre of the plot and there is lot of scope to decrease the fibre volume fraction and still satisfy the property constraint. Fig.\ref{Fig.12}b shows the microstructure generated by using the volume loss function in equation(19) and the scatter plot obtained at this volume fraction. The minimum fibre volume fraction obtained is 56.25\% and it can be observed from the plot that this configuration is one of the extreme cases produced for the given property setting. Using Equation 19, the theoretical volume fraction limits come out to be 33.33\% and 83.33\%, and our results lie well within these limits. Fig.\ref{Fig.13}b shows a similar case with different desired properties with a minimum fibre volume fraction resulting at 60.9\% in contrast to the most probable case at 67.2\% volume fraction in Figure 13a. The theoretical limits in this case lie at 33.33\% and 88.88\% . This proves that the inverse model can efficiently generate borderline cases at the extreme ends of the distribution by creating microstructures which may have rarely been seen before.\par\\ 
}
\section{Conclusion} \label{CONCLUSION}
In this work, we have presented a deep neural network-based predictive-generative architecture to design composite materials micro-structure for achieving the desired elastic properties. The deep neural network-based surrogate predictor model is trained from the data obtained through FEM analysis. To overcome the major challenge of the consumption of computational resources for the generation of the training data, we have presented a novel data augmentation scheme. The scheme increases the dataset by an order of magnitude and thus allows us to greatly reduce the computational costs incurred by the expensive FEM analysis. The data augmentation scheme is general and can be applied in any setting employing periodic boundary conditions. We then employed the developed property predictor model as a feedback mechanism in a generator model. The generator model is deployed for finding composite microstructures corresponding to desirable elastic response.

 We have also demonstrated that the generator model can incorporate volume fraction constraint and minimum volume fraction requirement in a robust manner and can generate desired microstructures in just a few seconds. The effectiveness and versatility of the generator model are demonstrated by the generation of microstructures in various settings. 

%\section{ Appendix}
%{\\Forward model: The forward model consists of five convolution layers and five dense layers. Each convolution layer has 64 filters of kernel size (3,3) and same padding. The first two dense layers have 256 nodes, the next two have 128 while the last dense layer is the output layer having three nodes for three elastic properties. The output layer has linear activation, while the rest of the layers have relu activation.
%\\Generator model: The input vector is a vector of 100 nodes each having randomly generated values. The generator dense layer has the same number of nodes as the number of pixels in the desired RVE. The layer is activated using sigmoid function. The time taken taken by the generator model to generate microstrucures in the various settings discussed is minimal, on the order of a few seconds.}



 \bibliographystyle{elsarticle-num} 
 \bibliography{ref}





% \end{thebibliography}
\end{document}